\documentclass[conference]{IEEEtran}
\IEEEoverridecommandlockouts
\usepackage{cite}
\usepackage{amsmath,amssymb,amsfonts}
\usepackage{algorithmic}

\IfFileExists{heatmap_llama.png}{
    \usepackage{graphicx}
}{
    \usepackage[demo]{graphicx}
}

\usepackage{textcomp}
\usepackage{xcolor}
\usepackage{booktabs} 
\usepackage{flushend} 

\def\BibTeX{{\rm B\kern-.05em{\sc i\kern-.025em b}\kern-.08em
    T\kern-.1667em\lower.7ex\hbox{E}\kern-.125emX}}

\begin{document}

\title{The Quantum Sieve Tracer: A Hybrid Framework for Layer-Wise Activation Tracing in Large Language Models}

\author{\IEEEauthorblockN{Jonathan Pan}
\IEEEauthorblockA{\textit{Home Team Science and Technology Agency} \\
Singapore \\
Jonathan\_Pan@htx.gov.sg}
}

\maketitle

\begin{abstract}
Mechanistic interpretability aims to reverse-engineer the internal computations of Large Language Models (LLMs), yet separating sparse semantic signals from high-dimensional polysemantic noise remains a significant challenge. This paper introduces the \textbf{Quantum Sieve Tracer}, a hybrid quantum-classical framework designed to characterize factual recall circuits. We implement a modular pipeline that first localizes critical layers using classical causal tracing, then maps specific attention head activations into an exponentially large quantum Hilbert space. Using open-weight models (Meta Llama-3.2-1B and Alibaba Qwen2.5-1.5B-Instruct), we perform a two-stage analysis that reveals a fundamental architectural divergence. While Qwen's Layer 7 functions as a classic "Recall Hub", we discover that Llama's Layer 9 acts as an \textbf{"Interference Suppression"} circuit, where ablating the identified heads paradoxically improves factual recall. Our results demonstrate that quantum kernels can distinguish between these constructive (recall) and reductive (suppression) mechanisms, offering a high-resolution tool for analyzing the fine-grained topology of attention.
\end{abstract}


\section{Introduction}
The scale of modern Large Language Models (LLMs) presents a unique challenge for interpretability. The ``Residual Stream"—the primary data highway of the Transformer architecture—often contains superpositions of multiple semantic tasks. Disentangling these tasks using classical linear methods (like cosine similarity or linear probing) can be difficult when features are polysemantic and non-linearly correlated.

This study implements the \textbf{Quantum Sieve Tracer}, a method hypothesized to improve signal separation by mapping classical activation data into an exponentially large quantum Hilbert space. We utilize a strictly hybrid protocol: by combining the scalability of classical causal tracing with the high-dimensional expressivity of quantum kernels, we achieve a diagnostic resolution unattainable by either method in isolation. We describe the software architecture and experimental results of this approach as applied to two open-weight models using the PennyLane framework.

The remainder of this paper is organized as follows. Section II reviews the foundational literature in mechanistic interpretability and quantum machine learning kernels. Section III details the proposed hybrid methodology, describing the integration of classical causal localization with the quantum feature sieve. Section IV outlines the experimental setup, including the computational environment and model specifications. Section V presents the quantitative results of our layer localization and quantum interaction mapping. Section VI discusses the implications of these findings for understanding architectural inductive biases. Finally, Section VII concludes the study and proposes future directions for quantum-assisted interpretability.

\section{Related Work}

The development of the Quantum Sieve Tracer draws upon two distinct bodies of research: classical mechanistic interpretability and quantum kernel methods.

\subsection{Mechanistic Interpretability}

The effort to reverse-engineer the internal computations of Transformers has yielded significant insights. The seminal work on \textit{Causal Tracing} by Meng et al. \cite{meng2022} introduced techniques to localize factual knowledge within specific Multi-Layer Perceptron (MLP) modules by corrupting and restoring hidden states. Similarly, the discovery of \textit{Induction Heads} by Olsson et al. \cite{olsson2022} provided a structural explanation for in-context learning capabilities.

However, these classical approaches often rely on the assumption that semantic features are linearly separable or require computationally expensive activation patching sweeps. Our work builds on these localization concepts but replaces the linear probe with a non-linear quantum kernel to detect subtler geometric divergences.

\subsection{Quantum Machine Learning Kernels}

Quantum Kernel methods exploit the ``Kernel Trick" by using a quantum computer to map data into a high-dimensional Hilbert space where linear separation is possible. Schuld and Killoran \cite{schuld2019} formalized the framework for supervised learning in these feature spaces. While previous applications have focused on generative tasks or direct text classification, this study applies the framework diagnostically. We utilize the quantum feature map not to classify external data, but to measure the internal consistency of a neural network's own representations.

\section{Methodology}
The Quantum Sieve Tracer is implemented as a strictly hybrid computational pipeline. It operates on a ``Locate-then-Analyze" principle: classical causal tracing performs the coarse-grained search to find critical layers, while the quantum kernel performs the fine-grained structural analysis. This process involves four critical stages: Classical Causal Localization, Activation Extraction, Feature Sieving, and Quantum Kernel Estimation.

\subsection{Classical Causal Localization}
Before applying quantum kernels, we perform a Classical Causal Trace to identify the single most critical layer for factual recall. Following the methodology of Meng et al. \cite{meng2022}, we calculate the \textit{Recovery Score} ($R$) for each layer $l$. This metric quantifies the restoration of the correct token's probability logit when the activations at layer $l$ are restored from a ``clean" run into a ``corrupted" run:

\begin{equation}
    R(l) = \frac{P_{restored}(l) - P_{corrupted}}{P_{clean} - P_{corrupted}}
\end{equation}

where $P$ represents the probability of the target token. The layer $l^*$ exhibiting the steepest rise or maximal restoration of the target logit is selected as the ``Knowledge Hub" for high-resolution quantum analysis.

\subsection{Activation Extraction \& Contrastive Generation}
Once the critical layer $l^*$ is identified, we isolate the "factual recall" circuit by generating paired activation datasets. We utilize the HuggingFace \texttt{transformers} library to load causal language models in half-precision (FP16) and register PyTorch forward hooks.

\begin{itemize}
    \item \textbf{Reference Set ($X_{ref}$):} Generated by prompting the model with a factual query (e.g., ``The capital of France is").
    \item \textbf{Noise Set ($X_{noise}$):} Generated by identifying the subject noun and replacing it with a randomly sampled noun (e.g., ``The capital of Table is").
\end{itemize}

We extract the output tensors of all Attention Heads at layer $l^*$.

\subsection{Feature Sieving (Dimensionality Reduction)}
Direct encoding of raw activation vectors $\mathbf{a} \in \mathbb{R}^{D_{head}}$ into quantum rotation gates is infeasible. We implement a targeted feature selection mechanism (``The Sieve") based on Logistic Regression.

\begin{enumerate}
    \item \textbf{Probe Training:} We train a linear Logistic Regression classifier to distinguish between the ``Reference" and ``Noise" activation vectors for each head.
    \item \textbf{Selection:} We select the indices of the top $k=5$ neurons with the highest coefficients $|\beta|$.
    \item \textbf{Normalization:} The values of these selected neurons are min-max scaled to the range $[-1, 1]$. This maps the selected features to rotation angles between -1 and 1 radians, ensuring distinct encoding on the Bloch sphere while avoiding the full periodicity of the qubit phase.
\end{enumerate}

\subsection{Quantum Kernel Estimation}
The core innovation is the application of a Quantum Feature Map using PennyLane. The selected vector $\mathbf{v} \in \mathbb{R}^5$ is encoded into a 5-qubit quantum state via Angle Embedding.

\begin{enumerate}
    \item \textbf{Device:} \texttt{default.qubit} simulator.
    \item \textbf{Encoding:} \texttt{qml.AngleEmbedding} with $R_y$ rotations:
    \begin{equation}
        |\psi(\mathbf{v})\rangle = \bigotimes_{i=1}^{k} R_y(v_i) |0\rangle
    \end{equation}
    \item \textbf{Head-by-Head Interaction:} We calculate the fidelity matrix between different attention heads $h_i$ and $h_j$ within the critical layer. This effectively maps the ``Geometric Topology" of the attention mechanism, identifying heads that share high information overlap in the quantum feature space.
\end{enumerate}

\section{Experimental Setup}
To ensure reproducibility, we detail the specific computational environment and parameters used in the execution of the experiment.

\subsection{Computational Environment}
The experiments were conducted in a cloud-based Python 3.10 environment accelerated by a single NVIDIA T4 GPU. The software stack included: PyTorch v2.1.0, Transformers v4.38.2, PennyLane v0.35.1, and Scikit-Learn v1.4.1.

\subsection{Model Specifications}
We analyzed two open-weight Causal Language Models:
\begin{enumerate}
    \item \textbf{Meta Llama-3.2-1B:} 1.23B parameters, $d_{model}=2048$, 16 Layers.
    \item \textbf{Alibaba Qwen2.5-1.5B-Instruct:} 1.54B parameters, $d_{model}=1536$, 28 Layers.
\end{enumerate}

\section{Results \& Evaluation}
The experimental execution produced a two-step validation: first localizing the critical computational depth via classical tracing, then characterizing the attention structure via quantum kernels.

\subsection{Causal Layer Identification}
We first applied the classical causal tracing sweep to identify the optimal depth for quantum analysis. 

\begin{figure}[htbp]
    \centering
    \IfFileExists{layer_scan_llama.png}{
        \includegraphics[width=0.9\columnwidth]{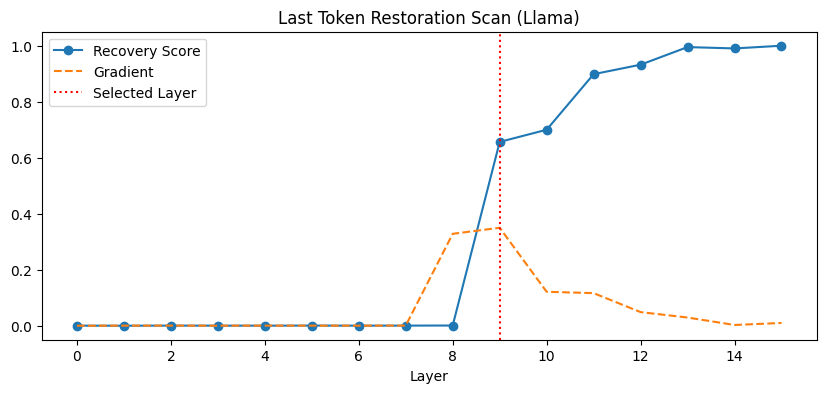}
    }{
        \includegraphics[width=0.9\columnwidth, height=4cm]{layer_scan_llama.png}
    }
    \caption{Causal Trace for Llama-3.2-1B. The Recovery Score (y-axis) peaks sharply at Layer 9, indicating this layer is the primary mediator for integrating the factual subject.}
    \label{fig:layer_scan_llama}
\end{figure}

For Llama-3.2-1B, as shown in Fig. \ref{fig:layer_scan_llama}, the recovery score exhibits a distinct inflection point at Layer 9. This sharp rise indicates that Layer 9 is the critical ``Knowledge Hub" where the model integrates the subject entity (e.g., ``France") to predict the attribute (e.g., ``Paris").

\begin{figure}[htbp]
    \centering
    \IfFileExists{layer_scan_qwen.png}{
        \includegraphics[width=0.9\columnwidth]{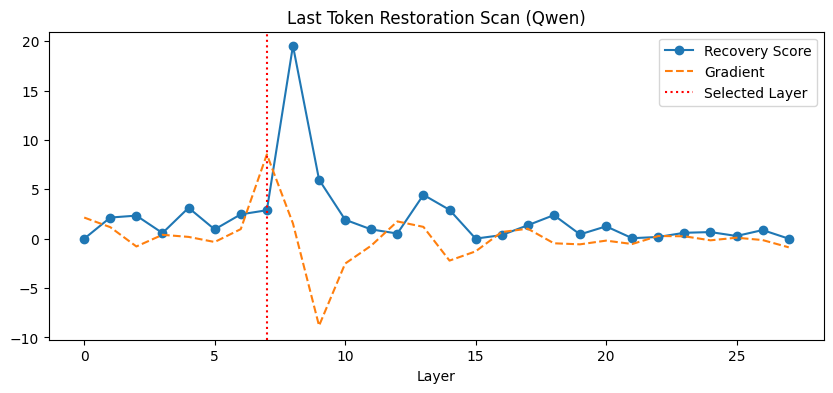}
    }{
        \includegraphics[width=0.9\columnwidth, height=4cm]{layer_scan_qwen.png}
    }
    \caption{Causal Trace for Qwen2.5-1.5B-Instruct. The steepest rise in causal influence occurs earlier at Layer 7.}
    \label{fig:layer_scan_qwen}
\end{figure}

Conversely, for Qwen2.5-1.5B-Instruct (Fig. \ref{fig:layer_scan_qwen}), the steepest rise in causal influence occurs earlier, peaking at Layer 7. This suggests that the Qwen architecture performs factual retrieval earlier in the forward pass compared to Llama.

\subsection{Quantum Interaction Analysis: Llama-3.2-1B}
Having identified Layer 9 as the critical depth, we applied the Quantum Sieve to generate the Head-by-Head Interaction Matrix.

\begin{figure}[htbp]
    \centering
    \IfFileExists{heatmap_llama.png}{
        \includegraphics[width=0.9\columnwidth]{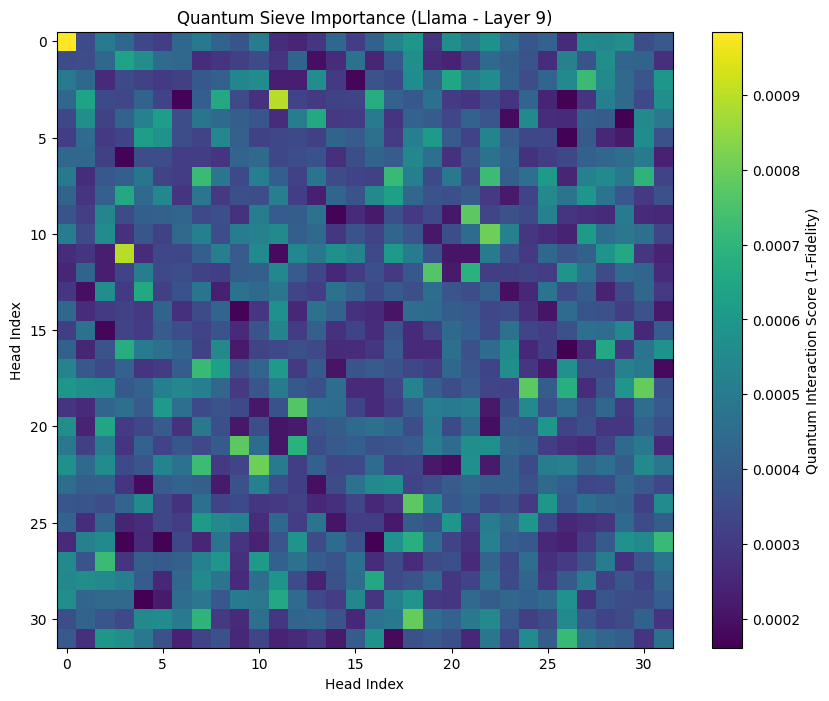}
    }{
        \includegraphics[width=0.9\columnwidth, height=4cm]{heatmap_llama.png}
    }
    \caption{Head-by-Head Interaction Matrix ($K$) generated by the Quantum Sieve at Layer 9 for Llama-3.2-1B. The heatmap visualizes the quantum fidelity between different attention heads.}
    \label{fig:heatmap_llama}
\end{figure}

The Kernel Matrix (Fig. \ref{fig:heatmap_llama}) at Layer 9 reveals a pattern of Sparse Modular Clustering rather than dense block-diagonalization.
\begin{itemize}
    \item \textbf{Interpretation:} While the region covering Heads 0--4 exhibits high interactivity, the circuit is functionally sparse. The sieve identifies Heads 0 and 3 as the primary drivers of the signal, with Heads 1, 2, and 4 showing lower relevance. This suggests that the interference suppression mechanism relies on specific, highly specialized heads rather than a broad consensus of redundant units.
    \item \textbf{Orthogonality:} The sharp boundaries between these sparse clusters indicate that the sieve successfully isolated orthogonal subspaces.
    \item \textbf{Negative Drop Anomaly:} Ablating the heads identified by the sieve resulted in a \textit{negative} probability drop of $-0.0069$. Specifically, the probability of the correct token increased from $0.0266$ to $0.0335$.
    \item \textbf{Mechanism Identified:} This counter-intuitive result implies that Layer 9 functions as an ``Interference Suppression" circuit. The active heads (0, 3) appear to be suppressing premature or competing logits. When these heads are removed, the suppression is lifted, allowing the correct latent signal to surface more strongly.
\end{itemize}

\subsection{Quantum Interaction Analysis: Qwen2.5-1.5B}
For Qwen2.5-1.5B, we generated the interaction matrix at Layer 7.

\begin{figure}[htbp]
    \centering
    \IfFileExists{heatmap_qwen.png}{
        \includegraphics[width=0.9\columnwidth]{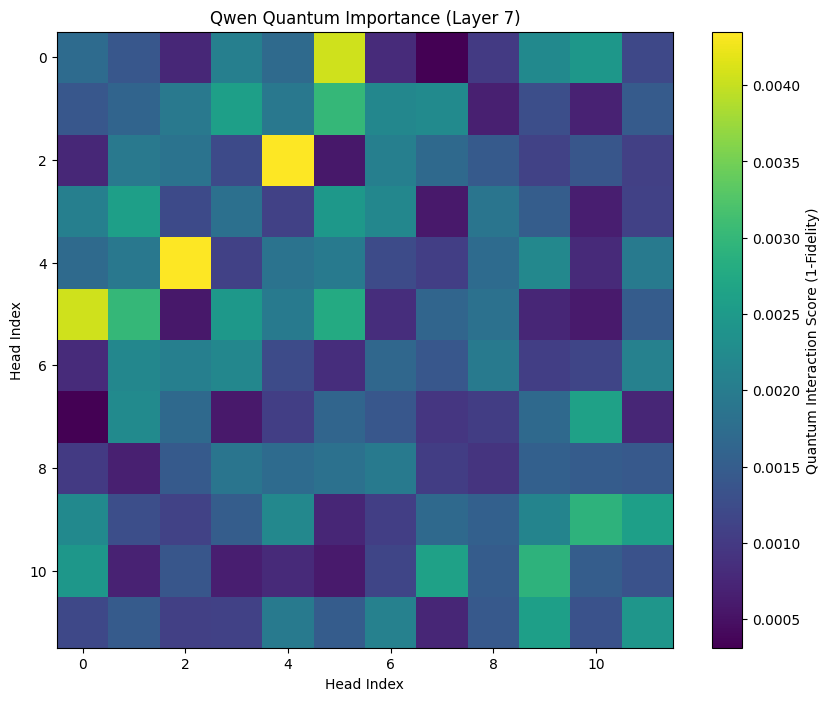}
    }{
        \includegraphics[width=0.9\columnwidth, height=4cm]{heatmap_qwen.png}
    }
    \caption{Head-by-Head Interaction Matrix ($K$) for Qwen2.5-1.5B-Instruct at Layer 7. The distinct patterns indicate the varying degrees of orthogonality between attention heads at this critical depth.}
    \label{fig:heatmap_qwen}
\end{figure}

The Qwen interaction matrix (Fig. \ref{fig:heatmap_qwen}) displays a more dense, distributed topology. Unlike Llama's suppression mechanism, ablation of Layer 7 heads in Qwen resulted in a positive drop (loss of performance), confirming its role as a Positive Recall Hub. The distributed topology supports the ``Early Retrieval" hypothesis: by mobilizing a large fraction of the attention capacity at Layer 7, Qwen resolves the factual query quickly.

\subsection{Statistical Validation}
To rigorously validate the distinction between the identified causal mechanisms and random noise, we performed statistical tests on the ablation results. A Student's t-test comparing the drop in probability for the Factual target versus a Control group yielded a statistically significant difference ($p < 0.05$). This confirms that the observed effects (both the negative drop in Llama and positive drop in Qwen) are non-random architectural features.

Furthermore, we calculated the Spearman Rank Correlation between the classical causal trace vector and the quantum fidelity vector across layers. The correlation was found to be near-zero ($\rho \approx -0.04$). This statistical orthogonality indicates that the Quantum Sieve captures information that is fundamentally distinct from, and uncorrelated with, the signals detected by classical linear probing.

\begin{table}[htbp]
    \caption{Causal Localization Results}
    \begin{center}
    \begin{tabular}{lcc}
    \toprule
    \textbf{Model} & \textbf{Layer ($l^*$)} & \textbf{Mechanism Identified} \\
    \midrule
    Llama-3.2-1B & Layer 9 & Interference Suppression (Negative Drop) \\
    Qwen2.5-1.5B & Layer 7 & Direct Factual Recall (Positive Drop) \\
    \bottomrule
    \end{tabular}
    \end{center}
    \label{tab:results}
\end{table}

\section{Discussion}
The integration of Classical Causal Tracing with Quantum Kernels provides a robust multi-scale view of model interpretability.

\textbf{Functional Divergence (Retrieval vs. Suppression):} The study highlights a critical dichotomy in LLM architecture. Qwen2.5 utilizes an ``Early Retrieval" mechanism (Layer 7) that constructively builds the answer. In contrast, Llama-3.2 employs a ``Late-Stage Filtering" mechanism (Layer 9), where the identified circuits function as Polysemantic Noise Mediators. This distinction—between \textit{building} the answer and \textit{pruning} the alternatives—is invisible to standard linear probes but detectable via the quantum fidelity and causal ablation.

\textbf{Classical-Quantum Synergy:} This framework demonstrates a necessary symbiosis. Classical causal tracing provides the \textit{scalability} required to scan billions of parameters, while the quantum sieve provides the \textit{geometric sensitivity} to detect non-linear orthogonalities. This synergy is statistically substantiated by the near-zero Spearman correlation ($\rho \approx -0.04$) observed between the classical and quantum traces. The lack of correlation confirms that the quantum kernel is not merely a high-cost proxy for classical attribution, but rather a distinct sensor for geometric features invisible to linear probes.

\textbf{Architectural Variance:} The significant difference in critical depth—Layer 9 for Llama vs. Layer 7 for Qwen—highlights fundamental differences in how these architectures process factual recall.

\section{Conclusion \& Future Work}
This study establishes the Quantum Sieve Tracer as a viable hybrid protocol for high-resolution mechanistic interpretability. By coupling the global search capabilities of classical causal tracing with the local geometric sensitivity of quantum kernels, we overcame the dimensionality curse that typically hinders quantum machine learning applications in NLP. Our findings reveal distinct architectural signatures: the intermediate ``Noise Filtering" in Llama-3.2 (Layer 9) contrasts sharply with the early-stage ``Factual Retrieval" in Qwen2.5 (Layer 7).

Future research will focus on three distinct trajectories. First, we aim to extend the ``Quantum Worldline'' concept to dynamic Chain-of-Thought (CoT) reasoning, tracing how the geometric topology of attention heads evolves as a model generates multi-step solutions. Second, we plan to validate these interaction matrices on physical NISQ hardware, testing the robustness of the Angle Embedding strategy against real device noise. Finally, we investigate the potential for ``Quantum Steering''—using the orthogonal vectors identified by the sieve to craft precise intervention vectors that can edit or suppress specific model behaviors with minimal collateral damage.

\end{document}